\documentclass[aps, prapplied, reprint, longbibliography]{revtex4-2}
\usepackage{xr-hyper, hyperref, graphicx, amsmath, amssymb, xcolor,lineno}
\graphicspath{{figures/}}

\begin{document}
\title{Photocurrent at oblique illumination and reconstruction of wavefront direction with 2d photodetectors}

\author{Kirill Kapralov$^1$, Vladislav Atlasov$^1$, Alina Khisameeva$^2$, Viacheslav Muravev$^3$, Weiwei Cai$^4$, Dmitry Svintsov$^1$}
\email[]{kapralov.kn@phystech.edu}
\affiliation{$^1$Moscow Institute of Physics and Technology, Dolgoprudny 141700, Russia}
\affiliation{$^2$Institute of Solid State Physics, Vienna University of Technology, 1040 Vienna, Austria}
\affiliation{$^3$Institute of Solid State Physics RAS, Chernogolovka 142432, Russia}
\affiliation{$^4$School of Mechanical Engineering, Shanghai Jiao Tong University, Shanghai, China}

\begin{abstract}
Many contemporary photodetectors operate beyond the readout of light intensity and enable the reconstruction of spectrum and polarization at the single-pixel level. However, the determination of light incidence direction with reconstructive detectors has not been realized so far. We show that photodetectors based on symmetric junctions of metals and 2d electron systems (2DES) enable (1) zero-bias photocurrent at oblique light incidence (2) reconstruction of incidence direction based on photocurrent measurements at variable carrier density. The former effect is based on peculiar electrodynamics of metal-contacted 2DES, where spatial variations of incident field phase translate into strong variations of local field amplitude. The local absorbances at two opposite metal-2DES junctions at oblique incidence are dissimilar, which results in finite photocurrent independent of microscopic rectification mechanism at these junctions. The direction of photocurrent uniquely determines the quadrant of light incidence. Quantitative determination of incidence angle becomes possible under conditions of 2d plasmon resonance at variable carrier density. In such a case, obliquely incident radiation excites the asymmetric plasmon modes, which amplitude carries unique information about angle of incidence.
\end{abstract}
\maketitle

\section{Introduction} There exists a series of electro-optical effects enabling the tuning of spectral and polarization photoresponse of semiconductors. These include quantum confined Stark effect~\cite{Harwit_Stark}, Frantz-Keldysh effect~\cite{Boer_FK_effect}, electric field effect translating into variable absorption~\cite{Allen1977}, to name a few. While these effects are observed since the middle of twentieth century, they were adopted only recently for the solution of an inverse problem, the reconstruction of spectrum and polarization via electrical tuning of photosensitive materials~\cite{Reconstructive1,Reconstructive2,Deep_otpical_sensing}. The field of reconstructive spectrometers boomed after demonstration of infrared spectroscopy with black phosphorus single-pixels~\cite{Yuan2021}, turned practical after invention of visible light spectrometers based on van der Waals junctions~\cite{Yoon2022}, and evolves toward spectroscopy with commercial III-V and organic semiconductor sensors~\cite{Organic_semicond_spectrometr,III_V_cascade_diode_spectrometer}. Reconstructive polarimetry takes its roots from electrically-controlled tensor of nonlinear conductivity in twisted bilayer graphene~\cite{Intelligent_infrared_TBG}, while recently it was extended to photodetectors using simple metal-graphene junctions~\cite{Semkin2026}. Independently, reconstructive spectroscopy and polarimetry with compacted combinations of multiple sensors, each having its own spectral and polarization response, achieved bright results~\cite{Full_Stokes_graphene,Belkin_2016,Deng_metasurface_array,Ganichev_ellipticity1}.

Despite advances of reconstructive polarimetry and spectroscopy, one extra degree of freedom for light still cannot be resolved in single-pixel sensors. This is the direction of propagation, parametrized by the wave vector ${\bf k}$. If realized, the reconstruction of wavefront direction may find applications for object tracking in automotive vision sensors. Detectors resolving the wavefront direction may determine the position of objects relative to the image plane, potentially enabling lensless imaging~\cite{lensless_imaging}. More generally, information about field phase is vital in microscopy, and in-sensor phase reconstruction would simplify the holographic imaging~\cite{Phase_Imaging}.

Intensity-based sensors cannot resolve the wave vector as it produces only the spatial variations of electric field phase. One microscopic mechanism of photoresponse, the photon drag~\cite{Pd_germanium,ganichev1983drag}, is sensitive to the direction of wave propagation and can be potentially used for wavefront direction reconstruction. Unfortunately, this effect is typically weak and masked by stronger photovoltaic and thermoelectric effects at the device contacts. The reason for photon drag smallness is the negligible magnitude of photon momentum, as compared to the thermal or Fermi momenta of charge carriers. In practical optoelectronics, the photon drag detectors have limited use for strong ultrafast laser pulses~\cite{PD_detectors}. Graphene-based detectors with asymmetric antenna arms were claimed sensitive to the phase difference of radiation fields at the arms, still, the effect was proved only for helicity determination~\cite{Matyushkin2020}.

\begin{figure*}[ht!]
    \includegraphics[width=1.0\linewidth]{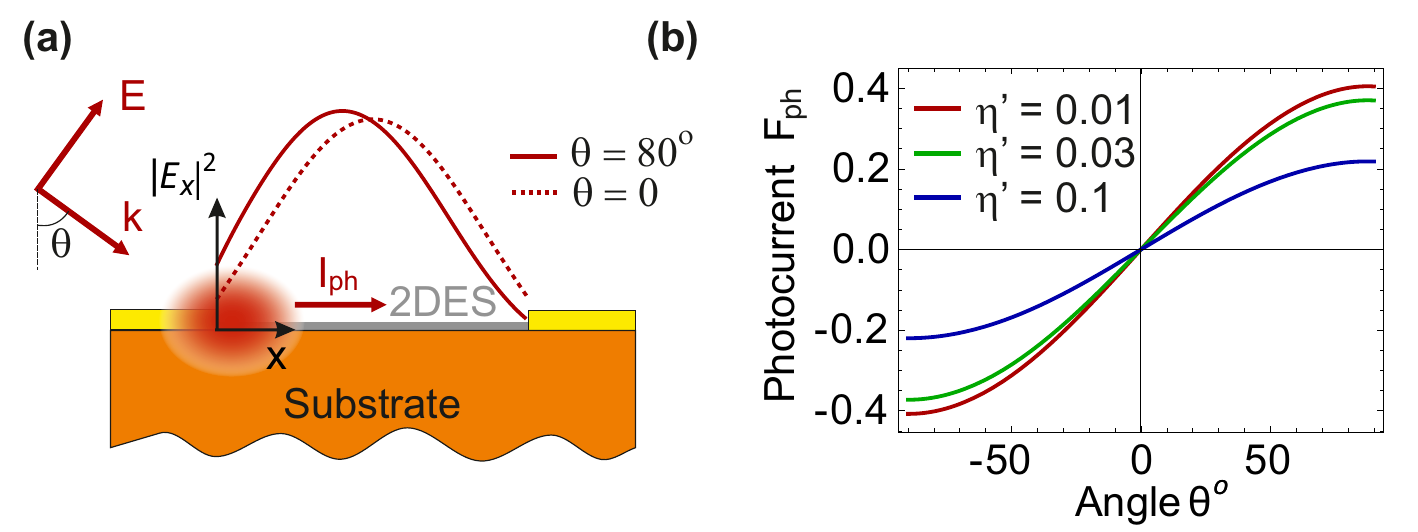}
    \caption{{\bf Zero-bias photocurrent in 2d photodetector at oblique incidence.} (a) Structure of metal-contacted 2d photodetector illuminated by obliquely incident radiation overlaid with profiles of local field $|E_x|^2$ at normal incidence (dashed) and inclined incidence (solid) (b) Computed photocurrent merit $F_{\rm ph}$ for subwavelength detector with $k_0L = 0.1$ at various values of normalized 2DES conductivity $\eta'$ (assumed real).}
    \label{fig1}
\end{figure*}

Here, we show that wave vector reconstruction is possible with intensity-based photodetectors based on metal-contacted 2d electron systems (2DES) and, in particular, 2d materials. Photocurrent generation in such devices is generally based on various rectification effects at the metal-semiconductor junctions in the zero-bias mode, and on the bulk photoconductivity~\cite{Amani_BP_Photoconductors} in the current-biased mode. All these effects are proportional to the local light intensity in the device plane and, on the first glance, lose the phase information. We find that the local light intensity depends in a non-trivial way on the phase of the incident light due to the diffraction at metal contacts and dynamic screening by the 2d channel itself. In other words, light scattering at the sensor itself turns phase modulation into amplitude modulation, which is further read out by conventional contact photoelectric effects.  

We find that the difference of local intensities at the two opposite contacts to the 2d channel (further called 'source' and 'drain') is proportional to $k_xL$, where $k_x = k_0\sin\theta$ is the projection of wave vector $k_0$ on the device plane, $\theta$ is the incidence angle and $L$ is the 2d channel length. This results in finite zero-bias photocurrent even if the source and drain junctions are identical, and the sign of this photocurrent is uniquely linked to the direction of propagation. We further find that quantitative determination of propagation direction is possible in the regime of 2d plasmon resonance in the photoconductive mode~\cite{Vasiliadou1993}. For oblique incidence, the spectrum of photoconductivity acquires an extra series of plasma peaks associated with excitation of spatially-odd modes. Relative magnitude of even and odd peaks is uniquely linked to the angle of incidence.

\section{ Detector structure and physics of photocurrent generation.} 

The structure under study represents a uniform 2DES of length $L$ with large metal contacts at each side, as shown in Fig.~\ref{fig1} (a).~\footnote{To track the conceptual idea, we consider metal contacts as semi-infinite, the width of the 2d channel is considered infinite as well.} The detector resides over a uniform substrate with permittivity $\varepsilon$. Light with $p$-polarized field ${\bf E}_0$ is incident at angle $\theta$.  Photocurrent generation in such structures typically occurs at metal-2DES contacts via photovoltaic~\cite{Echtermeyer2014} or photo-thermoelectic effects~\cite{Tielrooij-JPCM} (PVE and PTE), or via direct rectification at junction nonlinearities~\cite{Muravev2012_detector_spectrometer,Ryzhii_lateral_SJ}. 

Despite numerous possible rectification effects, their common property is the proportionality of photocurrent to the local density of wave electric field $|E|^2$. Denoting the local densities at the source and drain junctions as $|E_s|^2$ and $|E_d|^2$, and the microscopic responsivities of the respective junctions as $r_s$ and $r_d$, we can present the photocurrent as $I_{ph}= r_s |E_s|^2-r_d |E_d|^2$. Once $r_s$ and $r_d$ are dissimilar, e.g. due to the different composition of metals~\cite{Cai_sensitive_THz}, finite photocurrent appears upon uniform illumination. If the contacts are identical, $r_s = r_d \equiv r$, the photocurrent appears only for different local intensities at source and drain~\cite{Semkin_zero_bias}:
\begin{equation}
\label{eq-pv}
  I_{\rm ph}= r(|E_s|^2-|E_d|^2).
\end{equation}
We shall show that the difference in local intensities $|E_s|^2-|E_d|^2$ is nonzero for oblique incidence of radiation, which enables finite $I_{\rm ph}$ independent of rectification mechanism. 
To separate this purely electrodynamic effect of intensity asymmetrization from microscopic rectification physics, we introduce the dimensionless photocurrent figure of merit
\begin{equation}
F_{\rm ph} = \frac{|E_s|^2-|E_d|^2}{E_0^2}.
\end{equation}
The estimates of microscopic 'responsivities' $r$ for various rectification mechanisms can be found in Appendix A.

\section{ Model of electrodynamics}

We obtain the self-consistent field in the 2DES $E(x)$ via solution of combined Maxwell's equations and Ohm's law for high-frequency current density~\cite{Fateev_transformation,Tsymbalov_slot,Mikhailov_plasma}. The link between surface current density $J(x)$ and electric field $E(x)$ is obtained from the fundamental solution of the wave equation in the Lorentz gauge. It is most conveniently presented in the Fourier representation, where $q$ is the wave vector variable:
\begin{gather}
\label{eq-self-cons-Fourier}
Z_0 J\left( q \right)=g(q)\left[ E\left( q \right)-{\tilde{E}_{0}}\left( q \right) \right],\\
g(q)=i{{k}_{0}}\left( \frac{1}{\kappa_1(q)}+\frac{\varepsilon}{\kappa_\varepsilon(q)} \right),
\end{gather}
here $Z_0=377$ Ohm is the free-space impedance, $\kappa_\varepsilon \left( q \right)={{\left[ {{q}^{2}}-\varepsilon k_{0}^{2} \right]}^{1/2}}$ is the decay constant of electromagnetic field outside the 2DES, $\tilde E_0\left( q \right)$ is the Fourier transformed field at the top plane of the substrate {\it in the absence of 2DES}. It is related to the incident field $E_0$ as ${\tilde{E}_{0}}\left( q \right)=2\pi \delta \left( q-k_x \right){E}_{0}(1+r)\cos\theta$, where $r$ is the Fresnel's reflection coefficient from the substrate and $k_x = k_0\sin\theta$ is the $x$-projection of the wave vector.

Transition to the real-space formulation is achieved via inverse Fourier transform of (\ref{eq-self-cons-Fourier}) and results in~\footnote{We have intentionally presented the current $J(x)$ in Eq.~(\ref{eq-self-cons-coord}) as a convolution of electric fields at other positions $E(x')$ with the weight function $g(x-x')$. Such representation is convenient for objects enclosed in perfect conductors, the latter expelling the electric fields completely~\cite{Tsymbalov_slot}. As a result of perfect screening in the contacts, the integration in (\ref{eq-self-cons-coord}) is performed only along the length of 2DES, i.e. at $x\in[-L/2;L/2]$. Subsequent solution of integral equations in bounded domain is performed with relative ease. On the contrary, presenting the electric field $E(x)$ as a convolution of currents at distant points $J(x')$, we would encounter an integral equation on the infinite domain, which is problematic to solve.}
\begin{gather}
\label{eq-self-cons-coord}
Z_0J\left( x \right)=\int\limits_{-L/2}^{L/2} dx \ g\left( x-{x}' \right)E\left( {{x}'} \right)+2{{E}_{0}},\\
g\left( x-{x}' \right)=\int\limits_{-\infty}^{+\infty} \frac{dq}{2 \pi} \ {g(q){{e}^{i{{q}}\left( x-{x}' \right)}}}.
\end{gather}

The system (\ref{eq-self-cons-coord}) is supplemented by local Ohm's law, $J(x) = \sigma_{\rm 2d} E(x)$, where $\sigma_{\rm 2d}$ is the local conductivity of 2DES evaluated at the frequency of the incident field. The general method for solving integral equations of the form (\ref{eq-self-cons-coord}) lies in expansion of the field $E(x)$ into the orthogonal basis set~\cite{Tsymbalov_slot}, for which we choose the Legendre polynomials $P_m$:
\begin{equation}
\label{eq-Legendre}
E\left( x \right)={{E}_{0}}\sum\limits_{m=0}^{+\infty }{{{c}_{m}}{{P}_{m}}\left( 2x/L \right)}.  
\end{equation}

Introducing the representation (\ref{eq-Legendre}) into (\ref{eq-self-cons-coord}) and evaluating the integrals over $x$ and $x'$ explicitly, we find the matrix representation of the screening problem
\begin{gather}
\label{eq-self-cons-Legendre}
M_{km}c_m=i^k j_k\left(\frac{k_x L}{2}\right),\\
M_{km}=\frac{{{\delta }_{mk}}}{2k+1}\eta -\int\limits_{0}^{+\infty }{\frac{L dq}{4 \pi}  g(q)\operatorname{Re}\left[ {{\rho }_{k}}\left( \frac{qL}{2} \right)\rho _{m}^{*}\left( \frac{qL}{2} \right) \right]} 
\\
{{\rho }_{k}}\left( \frac{qL}{2} \right)=\int\limits_{-1}^{1}{{{e}^{i qL\xi/2 }}{{P}_{k}}\left( \xi  \right)d\xi }=2{{i}^{k}}{{j}_{k}}\left( \frac{qL}{2} \right),
\end{gather}
here $j_k$ are the spherical Bessel functions of the order $k$ and $\eta = \sigma_{\rm 2d}Z_0/2$ is the dimensionless 2d conductivity normalized by the free-space impedance. System (\ref{eq-self-cons-Legendre}) is solved by truncation of the screening matrix at finite size $k_{\max}$. The solution converges very rapidly, especially at low frequencies $\omega \lesssim c |\eta|/L$~\footnote{For very low-frequency field $\omega\rightarrow 0$, retention of a single Legendre polynomial $P_0 = 1$ is asymptotically exact, as one-dimensional current cannot be accumulated or lost in the dc limit, $J(x) = {\rm const}$.}. 


\section{ Main properties of photocurrent at oblique incidence}
Generation of photocurrent at oblique incidence is confirmed in Fig.~\ref{fig1} (b), where we show the computed merit $F_{\rm ph}$ as a function of the incidence angle. For deep-subwavelength size of the detector ($k_0 L = 0.1$ in Fig.~\ref{fig1} b) and purely real conductivity ($\eta = 0.01$), the angular dependence of photocurrent follows $F_{\rm ph}$ is odd with respect to $\theta$, tends to saturation at $\theta \rightarrow \pm \pi/2$, and tends to saturation with reduction in 2d conductivity $\eta'$ as well.  A closer inspection shows that, with a good accuracy, $F_{\rm ph} \propto \sin\theta$.

Explanation of these patterns can be obtained by decomposing the local electric field into spatially-even and spatially-odd components $E(x) = E_e(x)+E_o(x)$. Introducing the decomposition into expression for photocurrent merit (\ref{eq-pv}), we find
\begin{equation}
	\label{eq-photocurrent-odd-even}
	I_{\rm ph}= 4r {\rm Re}\left.[E_e(x) \times E^*_o(x)]\right|_{x=L/2}. 
\end{equation}
It implies that generation of photocurrent relies on the simultaneous excitation of odd and even modes. Further on, we can divide the Legendre polynomials used in our numerical scheme into even ($k=2K$) and odd ones ($k=2K+1$). The even and odd modes do not interact. Mathematically, the matrix of the screening equation (\ref{eq-self-cons-Legendre}) has zero elements for $m$ and $k$ of different parity. As a result, screening equations for even and odd $c$'s are solved independently. 

The odd field components are excited only for oblique incidence. Indeed, the driving terms for odd $k$ are proportional to $i^k j_k(k_xL/2)$, and all Bessel functions for $k\ge 1$ are zero for zero argument. Expanding the driving terms to the leading order in $k_xL/2$, we find that the odd field appears proportional to $i^1 j_1(k_xL/2) \approx i k_0 L \sin\theta / 2$. This explains the numerically observed $\sin\theta$- dependence of photocurrent.

The tilt-induced photocurrent is generally small, $F_{\rm ph} \ll 1$, as the variations of incident light phase along the subwavelength channel are tiny. Still, the dimensionless merit $F_{\rm ph}$ can greatly exceed the parameter $k_xL/2$ due to the concentration of the electric field in 2DES by perfectly conducting slit~\cite{Jadidi_MGM_plasmons}, $F_{\rm ph}\sim0.4$ in Fig.~\ref{fig1} (b) for $\eta'=0.01$ and for gliding incidence.

\begin{figure*}[ht]
    \includegraphics[width=1\linewidth]{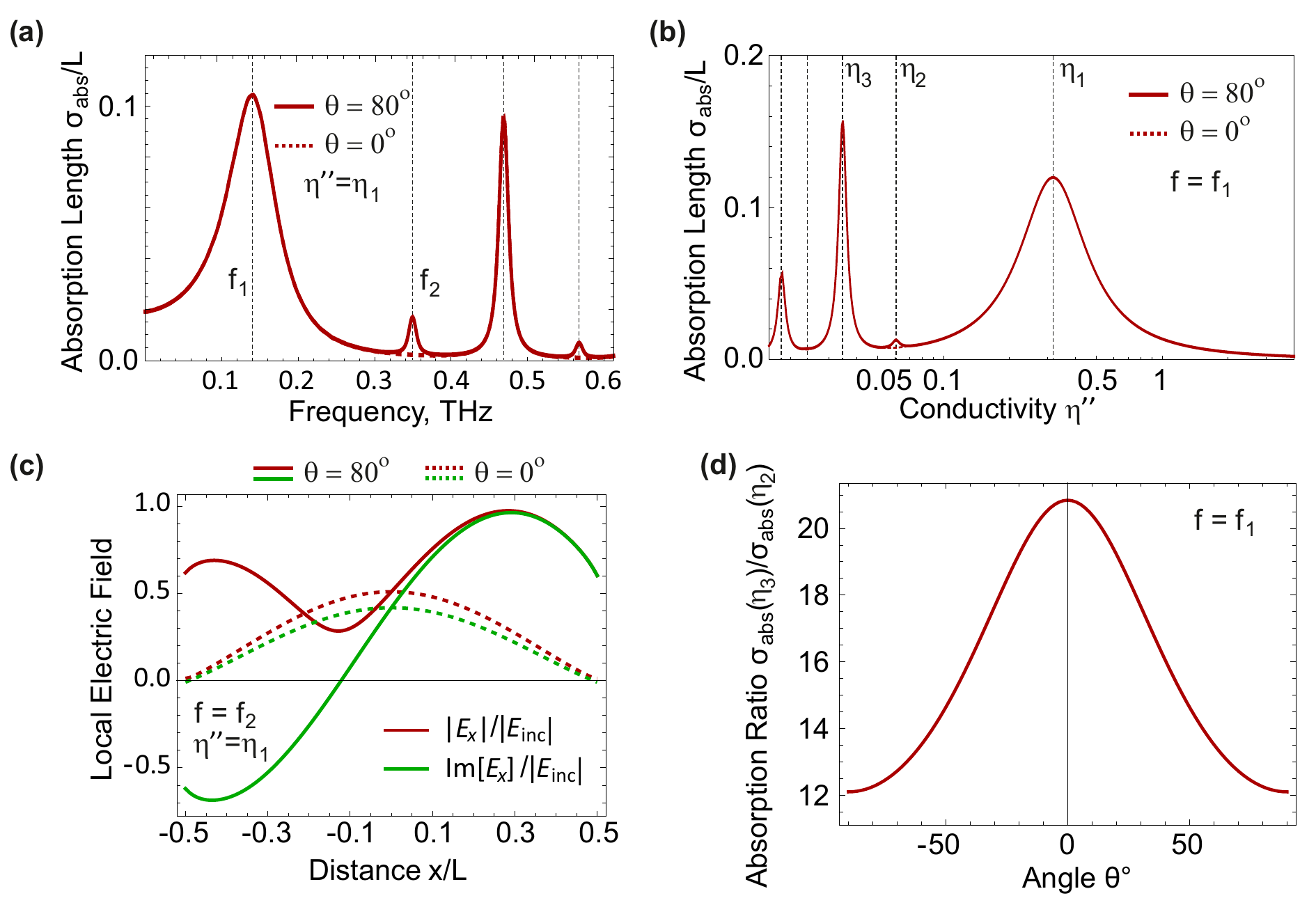}
    \caption{{\bf Direction-sensitive absorption in 2DES supporting plasmon resonance.} (a) Spectra of absorption length $\sigma_{\rm abs}/L$ for normal (dashed) and oblique (sold) incidence showing the emergence of extra dipole-forbidden resonances for inclined incidence (computed for $\eta''=0.31$). Vertical dashed lines mark the plasmon resonances. (b) Dependence of absorption length on the imaginary part of conductivity $\eta''$ for normal (dashed) and oblique (sold) incidence with the frequency of the incident field $f=0.14$ THz (c) Spatial profiles of local field $E(x)$ at the frequency of dipole-forbidden resonance for oblique (solid) and normal (dashed) incidence ($f=f_2=350$ GHz, $\eta''=0.31$) (d) the ratio of absorbances at the dipole-active ($\eta''=\eta_3$) and dipole-passive ($\eta'' = \eta_2$) resonances showed in panel (b) vs angle of incidence}
    \label{fig2}
\end{figure*}

The generation of photocurrent via screening-induced field asymmetry looks, from experimental viewpoint, very similar to the photon drag photocurrent. The latter is also proportional to the product of in-plane wave vector $k_x$ and the device length $L$~\cite{Karch_photon_drag_graphene}. Its magnitude, neglecting the screening effects, can be estimated as $I_{\rm pd} \approx (2\mu/\omega)  {\rm Re}\eta k_x |E|^2$, where $\mu$ is the carrier mobility. The ratio of $I_{\rm pd}$ and contact photocurrent induced by field asymmetry depends on the relation between $r$ and bulk drag responsivity. The contacts effects are definitely dominant at high frequencies and in samples with low mobility.

\section{Reconstruction of the incidence angle} 

The very presence of zero-bias photocurrent in symmetric 2d detectors at oblique light incidence is important for design of optoelectronic experiments. A more practical problem of reconstructive optoelectronics lies in resolution of the incidence angle via photocurrent measurements. Our calculations show that the direction of photocurrent is uniquely related to the sign of incidence angle (in a sub-wavelength device with $k_0L \ll 1$ and for real 2DES conductivity $\eta$). Pictorially, the contact glided by wave vector ${\bf k}$ acquires higher local electromagnetic density (source in Fig.~\ref{fig1}), the contact pierced by wave vector ${\bf k}$ acquires smaller $|E_x|^2$ (drain in Fig.~\ref{fig1}).

\begin{figure*}[ht!]
    \includegraphics[width=1\linewidth]{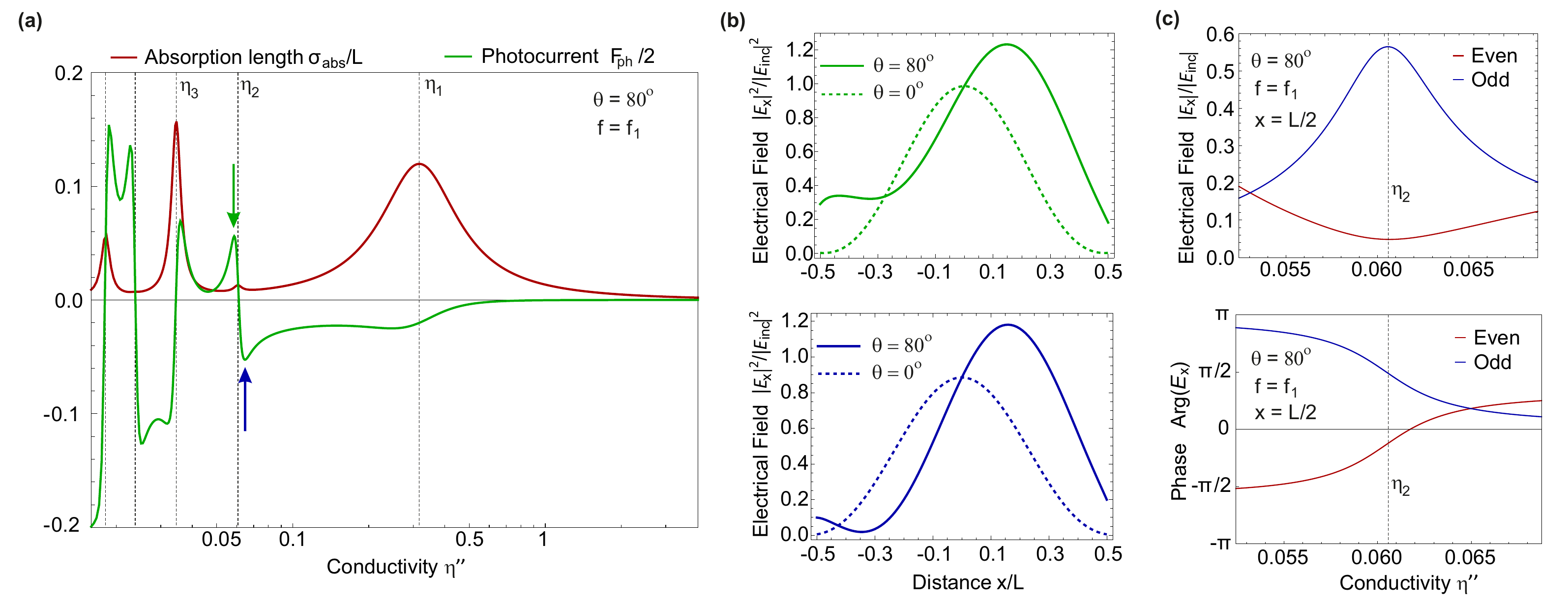}
    \caption{{\bf Zero-bias photocurrent at oblique incidence in 2DES supporting plasmon resonance.} (a) Dependences of absorption length $\sigma_{\rm abs}/L$ (red) and photocurrent merit $F_{\rm ph}$ (green) on the imaginary part of conductivity $\eta''$ at oblique incidence $\theta=80^\circ$ with $f=0.14$ THz (b) Spatial profiles of local field for selected values of conductivity marked by arrows in panel (a) (c) Magnitude (top) and phase $\mathrm{arg}[E(x=L/2)]$ (bottom) of the local electric field at $x=L/2$ in the vicinity of the dipole-forbidden resonance shown for the even (red) and odd (blue) field components. Calculation for $\theta=80^\circ$ and $f=0.14$ THz.}
    \label{fig3}
\end{figure*}

The quantitative determination of $\theta$, in line with the general procedures of reconstructive optoelectronics, can be based on sweep of detector parameters and subsequent processing of multiple photocurrents for these parameters. The convenient parameter is 2d conductivity $\eta$ controllable by gating. Unfortunately, a subwavelength detector with purely real conductivity $\eta$ is not suitable for quantitative reconstruction of $\theta$. The reason lies in a trivial dependence of photocurrent $I_{\rm ph} (\eta')$: independently on $\theta$, it grows toward saturation with reduction of $\eta'$. 

We proceed to show that quantitative angle reconstruction can be achieved with 2DES possessing complex conductivity $\eta = \eta'+i\eta''$, where $\eta''>0$ is the inductive part of conductivity. Under such conditions, the extended 2DES supports propagating 2d plasmons, while a confined 2DES-based detector supports plasmon resonances~\cite{Knap_resonant,Muravev2012_detector_spectrometer,Bandurin_resonant}. Large imaginary part of conductivity appears already at the level of free-electron Drude model, $\sigma_{\rm 2d} = ine^2/[m(\omega+i\tau)]$, with long momentum relaxation time $\omega\tau \gg 1$. The model is applicable to 2DES below the interband and intersubband absorption edges. In the present calculation, we use $L= 12\ \mu \text{m}$, $m = 0.24 m_e$ and $\tau = 0.25$ ps, corresponding to the high-quality 2DES at the GaN/AlGaN interface.

Excitation of plasmon resonance in confined 2DES occurs at resonant frequencies approximately satisfying 
\begin{equation}
\label{eq-pl-resonance}
\omega_k = c q_k \eta''(\omega),\qquad  q_k \sim \pi k/L,
\end{equation}
where $q_k$ is the quantized wave vector in the channel of length $L$ and $k$ is an integer. Resonant enhancement of local field results in elevated electromagnetic absorbance that can be read out electrically via photoconductivity~\cite{Vasiliadou1993}. For detectors with built-in asymmetry, plasmon resonance elevates the zero-bias photocurrent~\cite{Dyakonov_detection_mixing,Ryzhii_lateral_SJ,Knap_resonant,Muravev2012_detector_spectrometer,Bandurin_resonant}. We proceed to show that spectra of resonant absorption and photocurrent are essentially modified by oblique incidence.

The computed absorption length by 2DES $\sigma_{\rm abs}$ vs frequency $\omega/2\pi$ and conductivity $\eta''$ is shown in Figs.~\ref{fig2} (a) and (b), respectively (normalized by channel length $L$). The absorption at inclined incidence develops a palisade of resonances with alternating strong and weak peaks. Weak peaks emerge only at oblique incidence. Inspection of their electric field profiles reveals spatial asymmetry, i.e. the absence of average dipole moment. These dark (dipole-passive) modes require non-uniform incident field for excitation, with non-uniformity provided by the phase variations at oblique incidence. As the driving term for odd electric field is proportional to $k_xL$ in Eq.~\ref{eq-self-cons-Legendre}, the absorption at dipole-passive resonances would be proportional to $(k_x L)^2$. Measuring the ratio of absorbances at the dipole-passive and dipole-active resonances, one readily extracts $(k_x L)^2 = (k_0L)^2\sin^2\theta$ and determines the angle of incidence quantitatively. This algorithm is substantiated in Fig.~\ref{fig2} (d), where we plot the ratio of absorption lengths at even and odd resonances vs the angle of incidence. Finite offset of the curve from zero is due to the long shoulder of the fundamental resonance, and is readily subtracted upon conductivity sweep.

The zero-bias photocurrent $F_{\rm ph}$ as a function of conductivity $\eta''$ for 2DES supporting the plasmon resonance is shown in Fig.~\ref{fig3} (a) with green line. Contrary to absorption (marked by red line), it is not enhanced at resonant frequencies. Instead, it varies most strongly and changes sign at each resonance. Absolute maxima of $F_{\rm ph}$ are reached at both sides of absorption resonance. Profiling of local electric fields at photocurrent extrema shows that the degree of field asymmetry $|E_s|^2-|E_d|^2$ indeed changes across the resonance, Fig.~\ref{fig3} (b). The fundamental resonance at the largest $\eta_1$ is an exception: the photocurrent appears after the fundamental resonance, and approaches zero for very large $\eta$ when the channel field is uniform and plasmons are not excited.

To interpret this peculiar density-dependent photocurrent at oblique incidence, it is useful to recall the technique of phase demodulation known as complex phase shifting between carrier and signal. Harvesting finite photocurrent at oblique incidence is a similar problem: one transforms the phase variation of the external field into the difference of absolute fields at the source and drain $|E_s|^2-|E_d|^2$. The average incident field in our situation is interpreted as 'carrier', while the phase variation is the 'signal'. The driving terms for 'carrier' and 'signal' in self-consistent field equation (\ref{eq-self-cons-Legendre}) are $j_0(k_xL/2) \approx 1$ and $i j_1(k_xL/2) \approx i k_x L/2$. These terms are rotated by $\pi/2$ in the complex plane, which makes the intensity of {\it incident} field independent on $k_x$. The {\it local} field in 2DES is different from the incident by a complex-valued screening prefactor. Finite imaginary part of this prefactor (complex phase rotation) appears, eventually, due to the radiative and Ohmic losses. These result in non-Hermitian structure of the screening matrix $M_{km}$. For real screening prefactor, even and odd harmonics of the local field $E_e(x)$ and $E_o(x)$ would be $\pi/2$-shifted, and the photocurrent (\ref{eq-photocurrent-odd-even}) becomes zero.

With above preliminaries on phase demodulation, it becomes clear that the phase of local field $E_e(x)$ varies abruptly between 0 and $\pi$ upon passage of even (strong) resonances. This guarantees the intensity difference $|E_s|^2-|E_d|^2$ and finite photocurrent at the fundamental resonance. The phase of $E_o(x)$ varies abruptly between 0 and $\pi$ at weaker dipole-passive resonances (Fig.~\ref{fig3} c), which provides the sign-changing pattern of $F_{\rm ph}$. The absolute value of the product $E_e\times E_o$ is maximized at both even and odd resonances (Fig.~\ref{fig3} c), which results in photocurrent dips at non-resonant carrier densities. A similar picture of photocurrent was predicted very recently for 'plasmonic crystal', the 2DES with alternating regions of different carrier density, excited by the phase-modulated electromagnetic field~\cite{Gorbenko}. Our predicted photocurrent pattern at oblique incidence has much in common with flow resonances in magnetoplasmonics, the latter maximizing the magnitude of inverse Faraday effect~\cite{Flow_resonances}.

{\it Conclusion.} To conclude, we have theoretically demonstrated the feasibility of zero--bias photocurrent in symmetric metal-contacted 2d photodetectors under oblique illumination. The photocurrent appears due to the emergent asymmetry of local field intensity $E|(x)|^2$ within the channel and, hence, to the non-compensating rectified currents by the source and drain Schottky junctions. The effect has purely electrodynamic origin and persists for a range of electromagnetic frequencies and junction rectification mechanisms. For 2DES supporting the plasmon resonance, the density-dependent electromagnetic absorption develops a series of extra dipole-passive plasmon resonances. Measurement of their amplitude (e.g., via photoresistance) provides a tool for quantitative reconstruction of incidence angle from the photoresponse data. 

{\it Acknowledgement.} This work was supported by the grant No. 24-79-10081 of the Russian Science Foundation. 

\appendix
\section{Estimates of junction responsivities for different photocurrent generation mechanisms}
We present physical estimates of junction responsivities justifying the expression (\ref{eq-pv}) and providing microscopic estimates for responsivity $r$.

The photo-thermoelectric voltage is proportional to the difference of Seebeck coefficients for metal ($S_M$) and 2DES ($S_{2DES}$), and to the difference of hot electron temperatures at the source ($\Delta {{T}_{S}}$) and drain ($\Delta {{T}_{D}}$) junctions:
\begin{equation}
{{V}_{\rm PTE}}=\left( {{S}_{M}}-{{S}_{2DES}} \right)\left( \Delta {{T}_{S}}-\Delta {{T}_{D}} \right).
\end{equation}
The photocurrent is obtained by dividing the photovoltage by total device resistance, including two junction resistances ($2R_J$) and the resistance of 2DES itself (${R}_{2DES}$):
\begin{equation}
{{I}_{ph}}=\frac{{{S}_{M}}-{{S}_{2DES}}}{{{R}_{2DES}}+2{{R}_{J}}}\left( \Delta {{T}_{S}}-\Delta {{T}_{D}} \right).
\end{equation}
The electron overheating can be obtained from heat balance between Joule heating by radiation ($q_J(x) = \frac{1}{2}\operatorname{Re}{{\sigma }_{2D}}{{\left| {{E}(x)} \right|}^{2}}$) and heat loss into substrate phonons with characteristic time $\tau_E$:
\begin{equation}\frac{{{C}_{A}}}{{{\tau }_{E}}}\Delta {{T}_{S/D}}=\frac{1}{2}\operatorname{Re}{{\sigma }_{2D}}{{\left| {{E}_{S/D}} \right|}^{2}},
\end{equation}
here $C_A$ is the heat capacitance of electronic subsystem in 2DES per unit area $A$. Combining the above expressions, we arrive at the form (\ref{eq-pv}) with the established prefactor:
\begin{equation}
{{I}_{ph}}=\frac{1}{2}\operatorname{Re}{{\sigma }_{2D}}\frac{{{\tau }_{E}}}{{{C}_{A}}}\frac{{{S}_{M}}-{{S}_{2DES}}}{{{R}_{2DES}}+2{{R}_{J}}}\left( {{\left| {{E}_{S}} \right|}^{2}}-{{\left| {{E}_{D}} \right|}^{2}} \right).
\end{equation}

To estimate the photocurrent via direct rectification at the Schottky junction nonlinearity, we assume that the nonlinear current-voltage characteristic of each junction $I_J(V)$ is known. The rectified current at an individual junction for oscillating voltage $V_{S/D}(t)$ is obtained by Taylor expansion of the $I_J(V)$-curve with subsequent time averaging:
\begin{equation}I_{ph}^{S/D}=\frac{1}{2}\frac{{{d}^{2}}{{I}_{J}}}{d{{V}^{2}}}{{\left\langle {{V^2_{S/D}(t) }^{2}} \right\rangle }_{T}}
\end{equation}
The voltage generated by each junction is obtained by multiplying the photocurrent by junction resistance $R_J$. Voltage drop at the junction can be estimated as electric field timed by the junction length $l_J$. This leads us to the photovoltage estimate:
\begin{equation}V_{ph}^{S/D}=\frac{{{R}_{J}}}{4}\frac{{{d}^{2}}{{I}_{J}}}{d{{V}^{2}}}l_{J}^{2}{{\left| E_{S/D}^{{}} \right|}^{2}}
\end{equation}
Combining the two junction photovoltages and dividing by the total resistance of the structure, we arrive at the global photcurrent of the form:
\begin{equation}
{{I}_{ph}}=\frac{1}{4}\frac{{{R}_{J}}}{2{{R}_{J}}+{{R}_{2DES}}}\frac{{{d}^{2}}{{I}_{J}}}{d{{V}^{2}}}l_{J}^{2}\left( {{\left| {{E}_{S}} \right|}^{2}}-{{\left| {{E}_{D}} \right|}^{2}} \right)
\end{equation}

Finally, we estimate the photocurrent due to electron-hole separation by the built-in field of the Schottky barrier. With the neglect of recombination (for junctions of length $l_J<\sqrt{D\tau_R}$), each absorbed photon produces one electron in the external circuit. The number of absorbed photons equals the absorbed energy $q_J\times l_J$ divided by the quantum energy $\hbar\omega$. This results in the photocurrent of the individual junction of the form:
\begin{equation}
I_{ph}^{S/D}=\left| e \right|\frac{\frac{1}{2}\operatorname{Re}{{\sigma }_{2d}}{{\left| {{E}_{S/D}} \right|}^{2}}{{l}_{J}}}{\hbar \omega }.
\end{equation}
Photocurrent produced by the two junctions is obtained, as before, from an equivalent circuit comprising two junction resistances $2R_J$ in series with 2DES:
\begin{equation}
{{I}_{ph}}=\frac{\left| e \right|}{2}\frac{{{R}_{J}}}{2{{R}_{J}}+{{R}_{2DES}}}\frac{\operatorname{Re}{{\sigma }_{2d}}{{l}_{J}}}{\hbar \omega }\left( {{\left| {{E}_{S}} \right|}^{2}}-{{\left| {{E}_{D}} \right|}^{2}} \right).
\end{equation}
To conclude, we observe that the photocurrent produced via various junction rectification mechanisms is proportional to the difference of squared local fields at these junctions.

\bibliography{references}

\end{document}